\documentclass[superscriptaddress,floatfix,prl,onecolumn,amsmath,amssymb]{wlscirep}

\usepackage{graphicx}
\usepackage{dcolumn}
\usepackage{bm}
\usepackage{amssymb}
\usepackage{amsfonts}
\usepackage{amsmath}
\usepackage{mathtools}
\usepackage{color}
\usepackage{datetime}
\usepackage{footnote}
\usepackage{bbold}
\usepackage[normalem]{ulem}
 \usepackage{ragged2e}
 \usepackage{xfrac}
 \usepackage{sidecap}
 \usepackage{setspace}
 \usepackage{multirow}
 \usepackage{xcolor}
 \usepackage{flushend}
\usepackage{hyperref}
\usepackage{cleveref}
\usepackage{microtype}
\usepackage{siunitx}
\usepackage{etoolbox}
\usepackage{tikz}
\usepackage{makecell}
\usepackage{booktabs}
\usepackage{braket}
\usepackage[numbers,super]{natbib}

\newcommand{\gray}[1] {\textcolor{lightgray}{#1}}
\newrobustcmd*{\mysquare}[1]{\tikz{\filldraw[draw=#1,fill=#1] (0,0) rectangle (0.2cm,0.2cm);}}

\graphicspath{ {./..} }

\begin{document}

\author[1,$\dagger$]{Tim F. Weiss}
\author[1]{Hamed Arianfard}
\author[1]{Yang Yang}
\author[1,2,*]{Alberto Peruzzo}
\affil[1]{Quantum Photonics Laboratory and Centre for Quantum Computation and Communication Technology, RMIT University, Melbourne, VIC 3000, Australia}
\affil[2]{Qubit Pharmaceuticals, Advanced Research Department, Paris, France}

\affil[$\dagger$]{timweiss001@gmail.com}
\affil[*]{alberto.peruzzo@gmail.com}

\title{Spontaneous Parametric Down Conversion without poling for Silicon Carbide and Lithium Niobate photonics}

\begin{abstract}
State-of-the-art photon sources based on spontaneous parametric down-conversion (SPDC) currently rely on artificial structuring of the material nonlinearity to satisfy phase-matching conditions. This technique, known as periodic poling, is available only in a limited number of material platforms and introduces additional fabrication steps and errors, which are detrimental to up-scaling efforts.   
Here, we present a device architecture that enables SPDC of a wide range of frequencies without the need for periodic poling. We present explicit designs and calculations for 4H Silicon Carbide on-insulator, in which SPDC photon generation is so far unavailable, and thin-film Lithium Niobate on-insulator, a state-of-the-art quantum photonics platform. Our design, based on mode conversion and subsequent modal phase-matched SPDC, facilitates a CMOS compatible $\chi^{(2)}$ platform, and simplifies photon sources by removing the requirement of periodic poling and the associated additional fabrication complexity.
\end{abstract}



\maketitle

\section{Introduction}
(Single)-photon sources represent integral components in almost every optical quantum technology, including communication, computation and metrology \cite{Gisin:07,O'Brian:07,Giovannetti:11}. Intrinsically flexible and exceptionally engineerable, photon sources based on spontaneous parametric down-conversion (SPDC) have long represented the gold standard for many such applications \cite{Hong:87,Giustina:15,Yin:20,Wang:21}.

State-of-the-art integrated photonics SPDC photon sources generally rely on (periodic) poling of the material non-linearity, known as quasi-phase-matching (QPM), which is needed to satisfy momentum conservation, crucial for efficient photon generation \cite{Armstrong:62,Fejer:92}. While this technique allows for almost arbitrary, highly efficient down-conversion, it is available only in a limited number of material platforms. For instance, photonic integrated circuits based on Silicon Carbide \cite{Castelletto:22}, a promising emerging platform for integrated quantum technologies, have currently no way of realizing SPDC photon generation.

QPM requires additional fabrication procedures \cite{Shur:15}, which introduces a layer of fabrication errors, increases overall fabrication complexity, and may ultimately stand in the way of technological up-scaling. The availability of poling-free SPDC in a state-of-the-art quantum photonics platform like Lithium Niobate on-insulator\cite{BoesMitchell:23,Saravi:21} could considerably improve the availability of robust, integrated photon sources.

In this paper we present an alternative photon source design, relying on modal-phase-matching, allowing straightforward application for heralding, one- or two-photon interference. This effectively eliminates the fabrication-steps, the hardware and the errors associated with the poling part of the fabrication process, and enables SPDC in platforms where it could previously not be realized. To this end, we present explicit phase-matching calculations based on state-of-the-art integrated waveguides (WGs) in 4H silicon carbide (4H-SiC) on-insulator and Lithium Niobate (LN) on-insulator. For both platforms, the phase-matching conditions are tolerant to deviations from the designed WG geometry due to fabrication errors.

Modal-phase-matching already represents a well established technique in classical nonlinear optics, and has been used demonstrate to frequency conversion by matching incident fundamental modes to higher order waveguide modes \cite{LuoLin:18,LuoLin:19}. The technique has, however, so far not been adapted for photon generation processes via SPDC, due to the difficulty in generating the higher-order waveguide modes, here required initially, and the multitude of often very nuanced phase-matching processes to consider. Here, we present an elegant, simple solution for the generation of the critical higher-order waveguide modes as well as their phase-matching together with a detailed study of the viable phase-matching processes, to reveal a wide range of possible generation processes.

\begin{figure*}
    \centering
    \includegraphics[width=1\columnwidth]{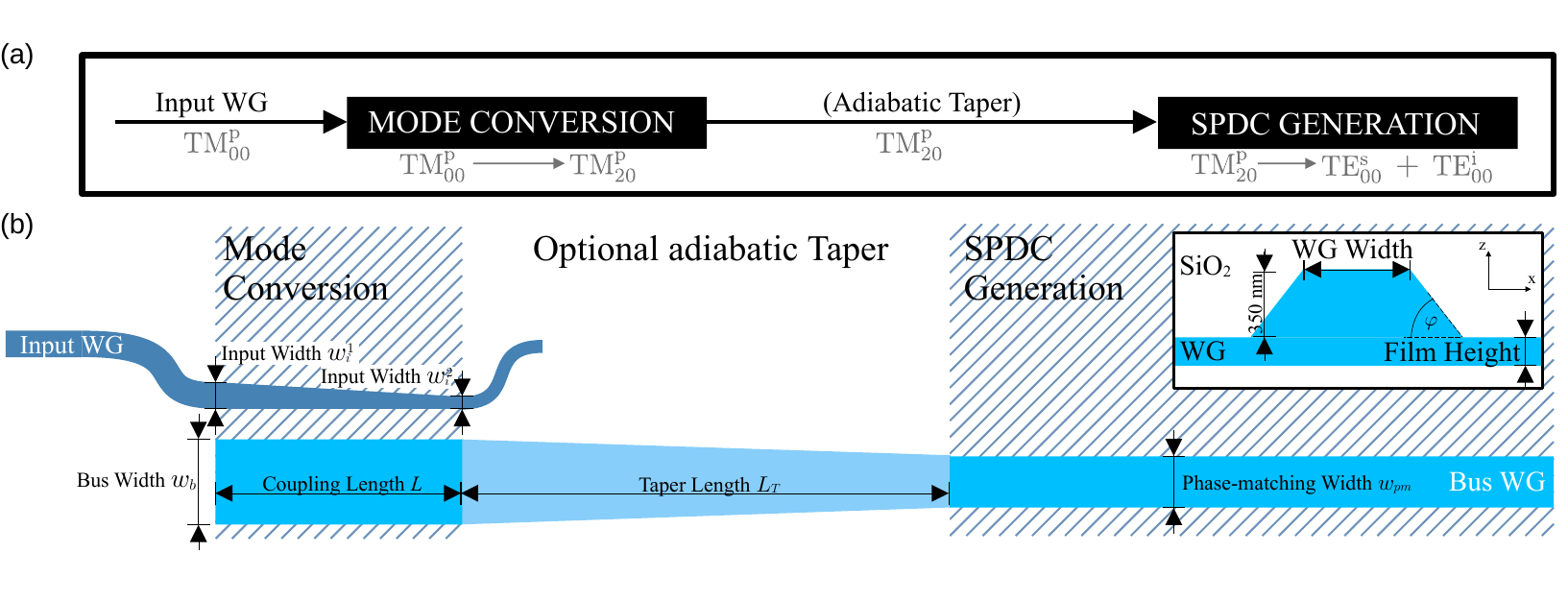}
    \caption{\textbf{Device design} \textbf{a}) Sketch of the device workflow, including conversion of (fiber-injected) $\mathrm{TM_{00}}$ modes into $\mathrm{TM_{20}}$ modes suited for down-conversion using an adiabatic directional coupler and SPDC  generation region, down-converting $\mathrm{TM_{20}}$ pump photons into $\mathrm{TE_{00}}$ signal and idler photons. An adiabatic taper may be included to relax WG geometry constraints of the mode converter. (\textbf{b}) Explicit design of the device described in (a) including relevant measurements. All WGs vary solely in their width, are implemented in z-cut crystals, and otherwise follow the design depicted in the inset. The phase-matching width is 650 nm and 1250 nm for the design in 4H-SiC on-insulator and LN on-insulator, respectively. The film-height and sidewall-angle $\varphi$ are 0 nm and 83\textdegree \: (150 nm and 60\textdegree) for 4H-SiC (LN).}
    \label{fig:Device}
\end{figure*}

\section{SPDC Phase-Matching}

Spontaneous parametric down-conversion is a three-wave-mixing process in which high-energy pump photons are down-converted into photon pairs of lower energy, generally referred to as signal and idler. To occur with optimal efficiency, the down-conversion process is required to satisfy strict conservation of energy and momentum, which is usually expressed via the phase-matching conditions \cite{Boyd:08}
\begin{eqnarray}
        \label{eq:EnergyConservation}
        \omega_{p} \; = \; \omega_{s} + \omega_{i} & & \begin{footnotesize} \textrm{\textbf{(energy conservation)}} \end{footnotesize} \\
        \label{eq:MomentumConservation}
        \beta_{p} \; = \; \beta_{s} + \beta_{i} & & \begin{footnotesize} \textrm{\textbf{(momentum conservation)}} \end{footnotesize},
\end{eqnarray}
wherein $\omega_{k}$, $\lambda_{k}$ and $\beta_{k}= (2 \pi) \frac{n^{\textrm{eff}}_{k}}{\lambda_{k}}$ refer to frequency, wavelength and propagation constants of the pump, signal and idler photons. The effective refractive indices $n^{\textrm{eff}}_{k}$ describe the photons' confinement as waveguide modes. The condition in eq. (\ref{eq:MomentumConservation}) is impossible to meet with fundamental modes in a conventional, uniform waveguide structure, as the shorter pump wavelengths possess a higher effective refractive index than both signal and idler, due to material dispersion and superior confinement within the waveguide.

\begin{figure*}
    \centering
    \includegraphics[width=1\columnwidth]{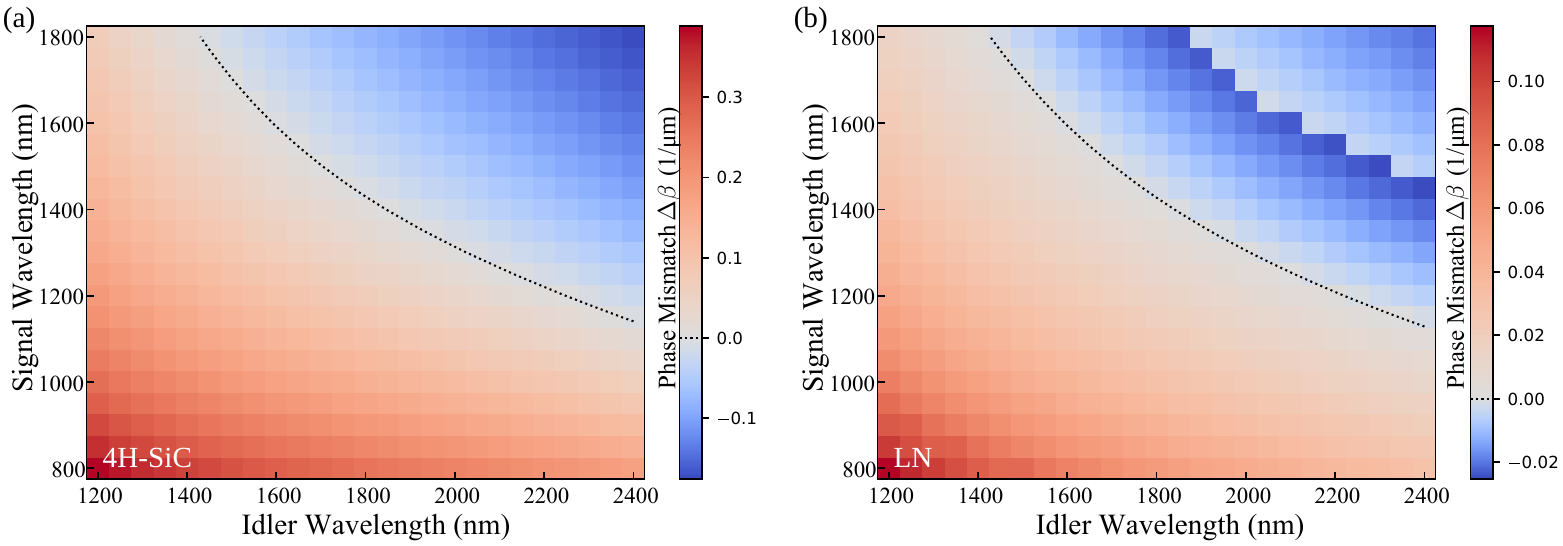}
    \caption{\textbf{SPDC phase-matching} Phase-mismatch $\Delta \beta = \beta^{\mathrm{TM_{20}}}_{p} - \beta^{\mathrm{TE_{00}}}_{s} - \beta^{\mathrm{TE_{00}}}_{i}$ in terms of the propagation constants for $\mathrm{TM_{20}}$ pump modes and $\mathrm{TE_{00}}$ signal/idler modes (neglecting factors $2 \pi$) for 4H-SiC on-insulator (\textbf{a}) and LN on-insulator (\textbf{b}). The dotted line highlights wavelength combinations with perfect phase-matching ($\Delta \beta = 0$), yielding maximum down-conversion efficiency. A specific pair of signal/idler wavelengths is selected via selecting the pump wavelength according to eq. (\ref{eq:EnergyConservation}). The discontinuity of the phase-matching condition in (b) is due to an anticrossing between the $\mathrm{TM_{20}}$ pump mode and the $\mathrm{TE_{30}}$ mode. We note that this phase-matching landscape is sensitive to the WG's geometry.}
    \label{fig:PhaseMatching}
\end{figure*}

To avoid this issue, we utilize phase-matching between second-order $\mathrm{TM_{20}}$ modes and fundamental $\mathrm{TE_{00}}$ modes in conventional z-cut ridge waveguides (see Fig. \ref{fig:Device}), effectively realizing type-I SPDC. The fully- and partially-etched waveguide geometries are adopted from state-of-the-art waveguide fabrication in 4H-SiC on-insulator and LN on-insulator, respectively. The waveguide etch-depth, film-height and sidewall-angle, in particular, are chosen to be representative of the most commonly employed and readily fabricated waveguides in both the 4H-SiC \cite{Lukin:20} and the LN \cite{Zhang:17,Lu:19,WuDiddams:24} platform. The waveguide width was adjusted manually, to optimize confinement of the $\mathrm{TM_{20}}$ mode while maintaining phase-matching over a wide range of wavelengths. We note that the $\mathrm{TM_{20}} \rightarrow \mathrm{TE_{00}}$ process is the only viable transition for this geometry, and present the phase-matching conditions of our device in Fig. \ref{fig:PhaseMatching}, revealing a wide range of available pump, signal and idler wavelength combinations.

Conversion between different order modes comes in hand with a reduction of the nonlinear modal overlap with respect to phase-matching between fundamental modes. A comparison of the expected down-conversion efficiency, given directly by the reduction of the modal overlap, for the transition considered here, a quasi-phase matched transition between fundamental modes and a hypothetical, naturally phase-matched process is depicted in Table \ref{tab:EfficiencyComparison}. We note, however, that the process involving fundamental modes is unavailable in 4H-SiC, while it relies on QPM in LN. Quantitatively calculating down-conversion efficiencies requires specifying experimental parameters beyond what is required for the functionality of the device proposed here, but can be done, in straight-forward fashion, along the theory outlined in the supplementary material.

\begin{table}[h!]
\centering
\begin{tabular}{ |p{1.3cm}|p{2.3cm}|p{1.9cm}|p{2.1cm}|  }

\multicolumn{1}{c}{} & \multicolumn{1}{c}{$\begin{subarray}{l} \text{\, naturally p.m.}\\ \text{$\mathrm{TM_{00}} \rightarrow \mathrm{TM_{00}}$} \end{subarray}$} & \multicolumn{1}{c}{$\begin{subarray}{l} \text{\; \; \; QPM}\\ \text{$\mathrm{TM_{00}} \rightarrow \mathrm{TM_{00}}$} \end{subarray}$} & \multicolumn{1}{c}{$\mathrm{TM_{20}} \rightarrow \mathrm{TE_{00}}$} \\
\hline
4H-SiC & \gray{100\%} & \gray{63.7\%} & 19.1\% \\
\hline
LN & \gray{100\%} & 63.7\% & 20.9\% \\
\hline
\multicolumn{4}{l}{\mysquare{lightgray}/\mysquare{black} {\footnotesize process unavailable/available in the respective platform}} \\
\end{tabular}
\caption{Down-conversion efficiencies calculated for the waveguide geometry in Fig. \ref{fig:Device}, for the $\mathrm{TM_{20}} \rightarrow \mathrm{TE_{00}}$ transition considered here and a quasi-phase matched transition between fundamental modes, calculated along the theory outlined in the supplementary material, in relation to a hypothetical, naturally phase-matched process (for an exemplary $800\, \mathrm{nm} \rightarrow 1600 \, \mathrm{nm}$ transition). We note, that a naturally phase-matched process between fundamental modes can not be realized in standard waveguides, and is listed as the ideal case for comparison only. Due to the different polarizations, the \text{$\mathrm{TM_{00}} \rightarrow \mathrm{TM_{00}}$} / \text{$\mathrm{TM_{20}} \rightarrow \mathrm{TE_{00}}$} transitions are mediated predominantly by the $d_{33}$/$d_{31}$ \cite{Nikogosyan:06,SatoKondo:09} elements of the nonlinearity; the efficiency reduction for the QPM approach is given by the Fourier coefficient associated with first order poling.}
\label{tab:EfficiencyComparison}
\end{table}

\section{Mode Conversion}

The phase-matching scheme described above relies on $\mathrm{TM_{20}}$ pump modes, which require an appropriate generation method. Here, we propose conversion of $\mathrm{TM_{00}}$ modes, which can be generated directly from fiber-to-chip coupling, into the required $\mathrm{TM_{20}}$ modes using an adiabatic directional coupler \cite{Zhao:23} (see Fig. \ref{fig:Device}). We note, that there are several other designs of mode converters, based on, for example, asymmetric directional couplers, Y-junctions, or multimode interference, that could be used instead \cite{Li:18}. The exemplary designs of the mode converters presented here were selected due to their relatively simple design, but may very well be adapted depending on application-specific demands or fabrication capabilities. The progressive inclusion of more adiabatic tapers, for instance, is associated with a trade-off between fabrication tolerance and component footprint \cite{Li:18}. Efficient mode converters have recently been fabricated in both the 4H-SiC on-insulator \cite{Shi:23} and the LN on-insulator platform \cite{Zhao:23}. 

\begin{figure*}
    \centering
    \includegraphics[width=1\columnwidth]{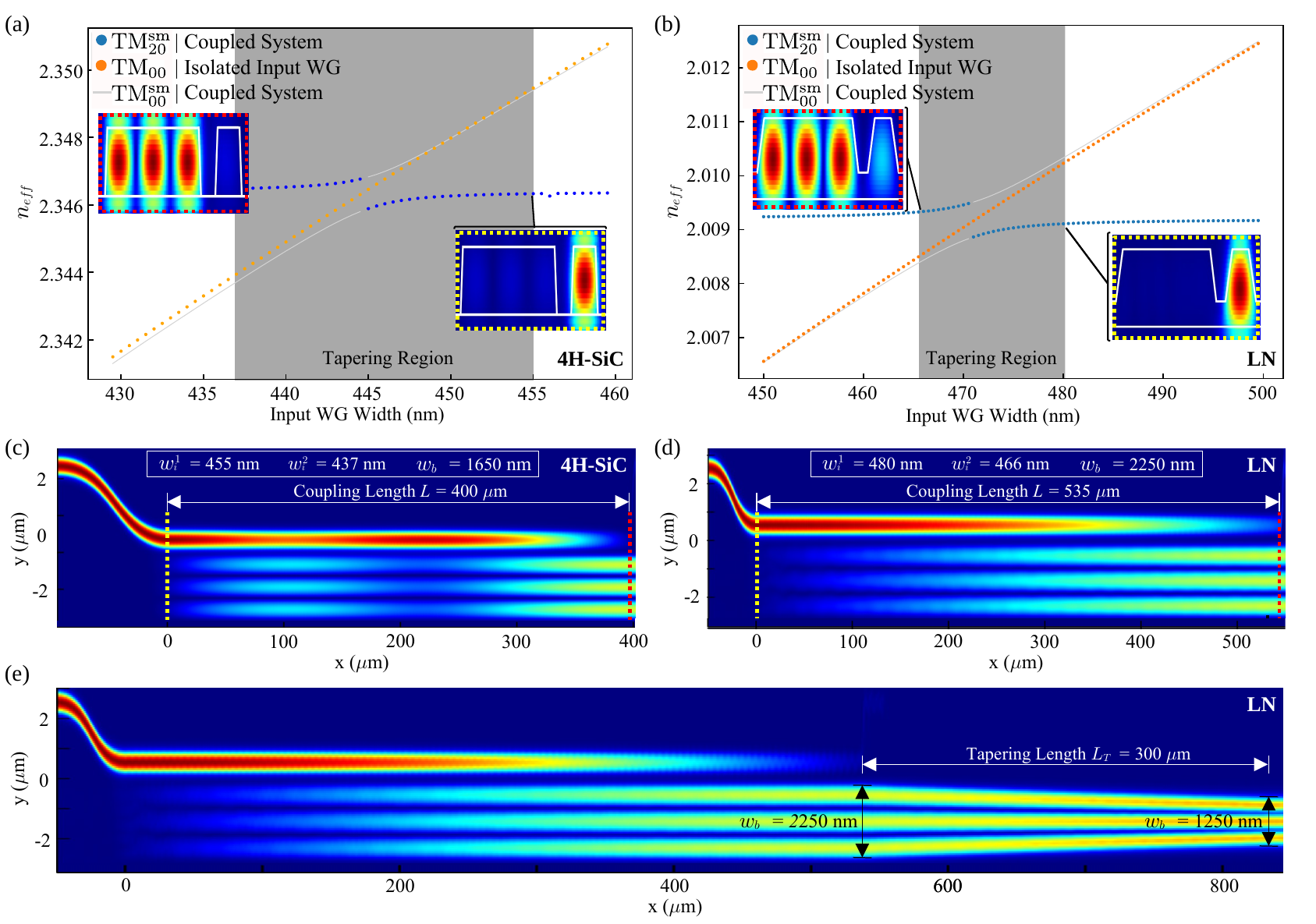}
    \caption{\textbf{Mode conversion} Effective refractive indices of the $\mathrm{TM_{20}^{sm}}$ supermode of the adiabatic directional-coupler based mode-converter (where there is a $\mathrm{TM_{20}}$ mode located in the bus WG and no light is present in the input WG, blue) and the $\mathrm{TM_{00}}$ mode of the isolated input WG (orange), plotted over the width of the input WG, for 4H-SiC on-insulator with a bus WG widths of 1650 nm (\textbf{a}) and LN on-insulator with a bus WG width of 2250 nm (\textbf{b}). The coupled system undergoes an anticrossing between the $\mathrm{TM_{20}^{sm}}$ supermode and the $\mathrm{TM_{00}^{sm}}$ supermode  (where a $\mathrm{TM_{00}}$ mode is located in the input WG and no light is present in the bus WG, indicated as a solid gray line for clarity) at the center of the conversion region. The taper is to sweep over the region of significant conversion until maximal conversion is achieved, as highlighted in gray. The insets depict the modes at the beginning and the end of the tapering region. FDTD simulations of the mode converters relying on a subsequent adiabatic taper for 4H-SiC on-insulator with bus WG width 1650 nm (\textbf{c}) and LN on-insulator with bus WG width 2250 nm (\textbf{d}) with conversion efficiencies of approx. 99\% and 96\%. (\textbf{e}) Example of the inclusion of an adiabatic taper used to relax the geometry constraints on the mode converter by subsequently tapering of the waveguide to the phase-matching width. The design parameters of the respective mode converters are depicted as insets in (c), (d). The gap between the two waveguides (defined as the separation between the WG ridges' bottom edges) in the FDTD simulations is 150 nm for 4H-SiC on-insulator and 200 nm for LN on-insulator.}
    \label{fig:ModeConversion}
\end{figure*}

The design of the mode-converter requires calculation of suitable WG geometries: the effective refractive index of the $\mathrm{TM_{20}^{sm}}$ supermode of the coupler (i.e. where there is a $\mathrm{TM_{20}}$ mode located in the bus WG and no light is present in the input WG) has to be equal to that of the $\mathrm{TM_{00}}$ mode of the isolated input WG at the center of the tapering region (see Fig. \ref{fig:ModeConversion}a,b)). The taper has to adiabatically (i.e. sufficiently slow) sweep over this condition. In our design, we achieve this by varying solely the WG top-width, which can easily be adapted to a fabrication mask. Once the widths are established, the total coupler length required for maximal conversion can be determined via a straight-forward propagation simulation.

We note that generating the $\mathrm{TM_{20}}$ mode directly in the WG phase-matched for SPDC demands a very thin input WG, which may lead to high losses and complications during fabrication. This can be avoided by designing a mode converter involving wider WGs, and subsequent adiabatic tapering of the bus WG to the width required for phase-matching. We present examples of the design including the adiabatic taper, exemplary for a pump wavelength at 800 nm, in Fig. \ref{fig:ModeConversion}, and report simulated conversion efficiencies of approx. 99\% and 96\% for 4H-SiC and LN respectively.


\section{Fabrication Tolerance and Tunability}
Due to the relative immaturity of SiC- and LN on-insulator WG fabrication, both platforms can show significant discrepancies between the fabricated WG geometry and the designed WG geometry. The main contributor to these deviations is the etching process, which can be modeled effectively via uncertainties in the etch-depth and the sidewall-angle of the WG. In this model, the WG bottom-width is given by the etching mask and remains constant, while the WG top-width changes dynamically. Accordingly, the gap between the two WGs of the mode converter (defined as the separation between the WG's bottom edge) remains unchanged.

Variations in the WG width due to writing errors in the etching mask can be reduced to below 5 nm utilizing precision Electron-Beam-Lithography, giving a negligible contribution.

\begin{figure*}
    \centering
    \includegraphics[width=1\columnwidth]{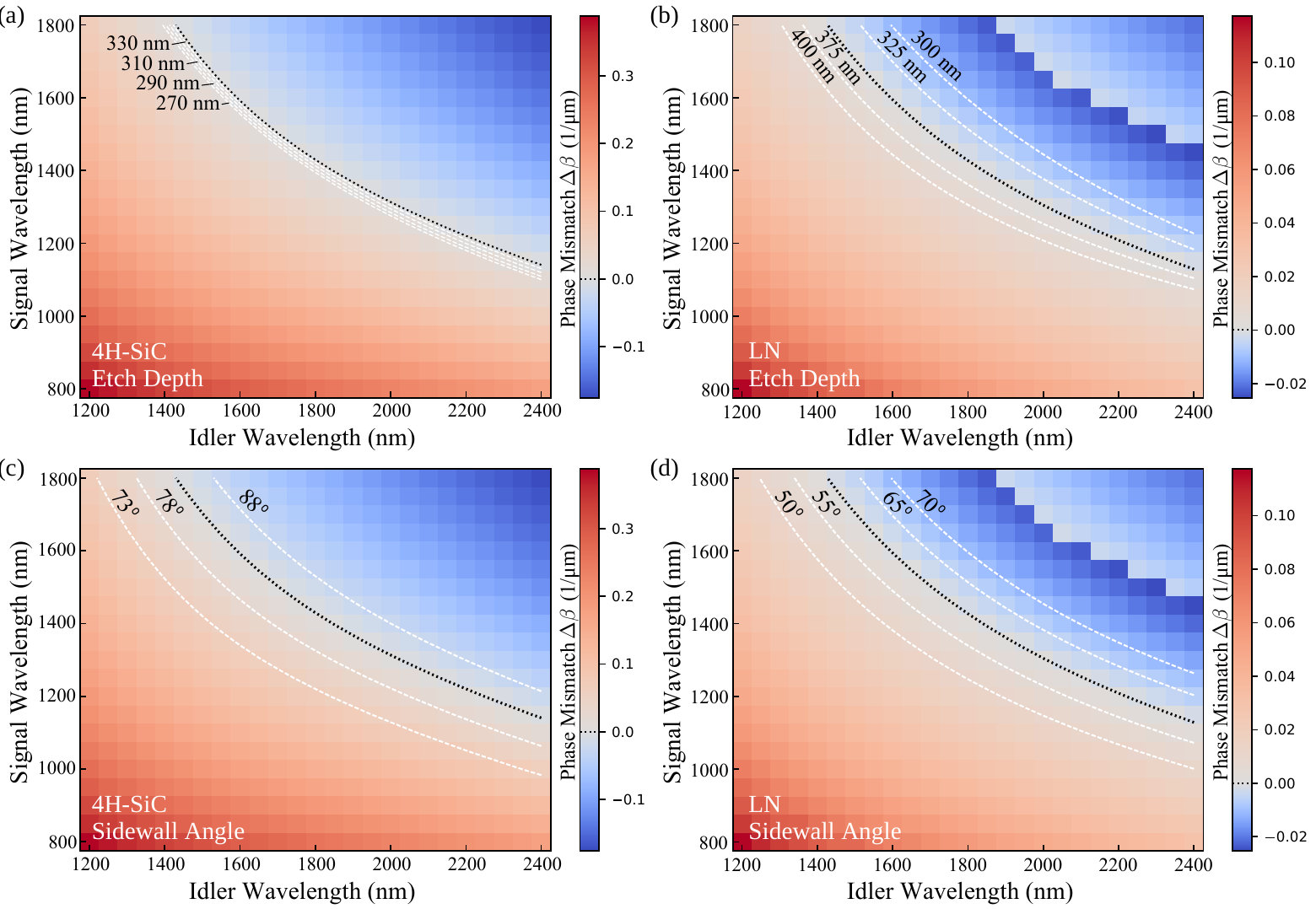}
    \caption{\textbf{Fabrication tolerance and tunability} Significance of WG geometry deviations (white lines) from the design (black line) during the fabrication process in terms of the etch-depth and the WG sidewall-angle, for 4H-SiC on-insulator (\textbf{a}), (\textbf{c}) and LN on-insulator (\textbf{b}), (\textbf{d}).The white lines hereby indicate the shifted pump/signal/idler wavelength combinations that correspond to perfect phase-matching. The black line and the background colormap correspond to the device fabricated perfectly to design. The efficiency drop-off associated with moving away from the lines of perfect phase-matching depends on both the effective length of the interaction region and the shape/bandwidth of the pump pulse, and requires case-specific calculation along the theory outlined in the supplementary material.}
    \label{fig:FabricationTolerance}
\end{figure*}

We present simulations capturing the effects of fabrication errors on the phase-matching conditions in Fig. \ref{fig:FabricationTolerance}.
We note, that if the fabrication process were to allow precise control over the waveguide geometry, this would instead allow to further tune the down-conversion process and extend the range of available phase-matching conditions.

The effective refractive indices required for SPDC phase-matching considerations were calculated using the mode-solver provided my COMSOL Multiphysics. Calculating the phase-matching curves presented in Figs. \ref{fig:PhaseMatching}, and \ref{fig:ModeConversion} follow directly from eqs. (\ref{eq:EnergyConservation}) and (\ref{eq:MomentumConservation}).  Propagation simulations were performed using FDTD simulations. The refractive indices of LN, 4H-SiC and $\mathrm{SiO_{2}}$ were adopted from Zelmon et al. \cite{Zelmon:97}, Wang et al. \cite{Wang:13} and Malitson \cite{Malitson:65} respectively.

\section{Discussion}

The proposed device architecture enables SPDC of a broad range of frequencies in conventional ridge WGs without the need for periodic poling. This makes (single)-photon generation via SPDC available in material platforms where QPM is not possible, and reduces fabrication complexity and -errors in platforms where it is. As a consequence, considering that other three-wave-mixing processes can be implemented in a similar fashion, 4H-SiC on-insulator may now be employed as a fully CMOS-compatible $\chi^{(2)}$ platform. 

We would like to, at this point, address explicitly the applicability of the proposed device for different utilizations of single-photon sources.

The photons generated by our scheme are identical to those generated by the corresponding QPM device, and can be utilized in the same manner. Accordingly, separation of non-degenerate photons, e.g. for photon heralding, can be achieved in a straightforward fashion by adding a wavelength selective directional coupler \cite{Krapick:13} or wavelength division multiplexer \cite{HeDai:24}. This may be extended to multiplexing architectures involving multiple heralded sources \cite{MeyerScott:20}.

When operated in the high-gain limit, the device may be used for the generation of squeezed states, in full analogy to sources of squeezed light based on simple QPM \cite{WuWu:86}.

The photons generated by the phase-matching scheme presented here are further immediately suitable for any application that aims to directly utilize the generated bi-photon state. Spectroscopy based on nonlinear interferometry \cite{Chekhova:16} or induced coherence for instance \cite{Chekhova:16,Kumar:20} requires entangled, often far non-degenerate photon pairs available over a broad range of frequencies. This can be achieved on a single device by tuning the pump wavelength \cite{Solntsev:18}. 

Applications relying on the interference of multiple (heralded) single photons from different sources achieve optimal fidelity only if the photons are in pure states. In 4H-SiC pure photons can be approximated via filtering (associated with a trade-off in photon counts and signal/idler number-correlations). In platforms where QPM is available, pure photon generation may be achieved optimally by variation of the material nonlinearity \cite{URen:06}, generally realized with non-(strictly)-periodic poling.

We note, that the presence of QPM does not make our phase-matching scheme redundant, but, on the contrary, allows to bypass certain key limitations associated with the poling technique. In particular backward-wave generation \cite{Harris:66,KuoCanalias:23}, currently unavailable in the integrated LN platform due to the minimum available inverted domain size, could be realized with a combination of modal- and QPM. Backward-wave generation may be used to generate narrowband photons for interfacing with solid-state memories \cite{LvovskyTittel:09,HeshamiSussman:16}, to facilitate group-velocity-matching essential for pure photon generation \cite{URen:06}, or to construct mirrorless optical-parametric-oscillators \cite{Canalias:07}.

Implementation of our device in LN on-insulator comes in hand with an intrinsic, notable reduction of the down-conversion rate with respect to those achievable with periodic poling due to decreased nonlinear modal overlap. Having said this, the actual photon-pair generation rate of the device is calculated from this baseline rate of down-conversion by selecting explicitly the pump power and the conversion region length. For perfect phase-matching, pair generation scales linearly and quadratically (or as $L^{\frac{3}{2}}$, if the entire generation spectrum is considered) with the pump power and the conversion region length $L$, respectively \cite{Helt:12}. We believe, that if these two parameters are chosen appropriately, the proposed device will prove to be applicable as a photon source in many applications. Device-length scaling, for instance, can be realized by appropriate spiral waveguide designs, which are commonly used in four-wave-mixing devices \cite{Pasquazi:10,Da:17} or via resonant confinement of the conversion region, which has already been realized for second-harmonic generation, leading to an efficiency increase much beyond the intrinsic reduction associated with this scheme \cite{Cheng:23,Lu:19}

The proposed device is, to the best of our knowledge, the first design aiming to achieve modal-phase-matched SPDC in integrated WGs. 

\smallskip

\noindent\textbf{Acknowledgments} AP acknowledges an RMIT University Vice-Chancellor’s Senior Research Fellowship and a Google Faculty Research Award. This work was supported by the Australian Government through the Australian Research Council under the Centre of Excellence scheme (No: CE170100012).

\smallskip

\noindent\textbf{Disclosures} The authors declare no conflicts of interest.

\smallskip

\noindent\textbf{Data availability} All data generated or analyzed during this study are included in this article.

%

\end{document}